# High thermoelectric power factor of poly(3-hexylthiophene) through in-plane alignment and doping with a molybdenum dithiolene complex

Viktoriia Untilova,[1] Jonna Hynynen,[2] Anna I. Hofmann,[2] Dorothea Scheunemann,[2] Yadong Zhang,[3] Stephen Barlow,[3] Martijn Kemerink,[4] Seth R. Marder,[3] Laure Biniek,[1] Christian Müller,[2]* Martin Brinkmann[1]*

[1] Université de Strasbourg, CNRS, ICS UPR 22, F-67000 Strasbourg, France

[2] Department of Chemistry and Chemical Engineering, Chalmers University of Technology, 41296 Göteborg, Sweden

[3] School of Chemistry & Biochemistry and Center for Organic Photonics and Electronics, Georgia Institute of Technology, Atlanta, Georgia 30332-0400, USA

[4] Centre of Advanced Materials, Heidelberg University, 69120 Heidelberg, Germany

e-mail: christian.muller@chalmers.se; martin.brinkmann@ics-cnrs.unistra.fr

TOC Graphic

**Abstract**

Here we report a record thermoelectric power factor of up to 160 μW m$^{-1}$ K$^{-2}$ for the conjugated polymer poly(3-hexylthiophene) (P3HT). This result is achieved through the combination of high-temperature rubbing of thin films together with the use of a large molybdenum dithiolene p-dopant with a high electron affinity. Comparison of the UV-vis-NIR spectra of the chemically doped samples to electrochemically oxidized material reveals an oxidation level of 10%, i.e. one polaron for every 10 repeat units. The high power factor arises due to an increase in the charge-carrier mobility and hence electrical conductivity along the rubbing direction. We conclude that P3HT, with its facile synthesis and outstanding processability, should not be ruled out as a potential thermoelectric material.

**Introduction**

Conjugated polymers currently receive considerable attention as thermoelectric materials because they are composed of abundant elements and offer ease of processing, low weight and mechanical robustness.[1-3] The thermoelectric efficacy of a material can be described by either the dimensionless figure of merit $ZT = \alpha^2 \sigma T / \kappa$ or the power factor $\alpha^2 \sigma$, where $\alpha$ is the Seebeck coefficient, $\sigma$ and $\kappa$ the electrical and thermal conductivity and $T$ the absolute temperature. Power factors reported for unoriented conjugated polymers now reach values of 10$^2$ μW m$^{-1}$ K$^{-2}$,[4-8] and in case of aligned materials more than 10$^3$ μW m$^{-1}$ K$^{-2}$.[9, 10] The majority of studies focus on polythiophenes because they are widely available and can be synthesized with a wide range of molecular weights, side-chain lengths and regioregularities. As a result, polythiophenes are ideal for elucidating fundamental structure-property relationships that underpin the thermoelectric performance of conjugated polymers. The archetypal semi-crystalline conjugated polymer is poly(3-hexylthiophene) (P3HT), which has been employed in a considerable number of studies related to organic thermoelectrics. Regardless, the highest thermoelectric power factor reported for P3HT has remained well below $10^2$ µW m$^{-1}$ K$^{-2}$ (see Table 1), as a result of which many researchers question the relevance of structure-property relationships established with this material.

The power factor of a wide range of organic materials approximately scales according to the empirical power law $\alpha^2\sigma \propto \sqrt{\sigma}$.[11-13] Hence, the majority of strategies for increasing the power factor concentrate on improving the electrical conductivity, which is given by:

$$\sigma = N_v \cdot \mu \cdot e \qquad (1)$$

where $N_v$ and $\mu$ are the density and mobility of charge carriers, and $e$ is the elementary charge.

Charges are created through molecular doping, i.e. dopant molecules are added to the semiconductor, which in case of p-doping take up an electron, leaving a hole behind. To reach a high conductivity it is paramount that the fraction of mobile charges created per dopant molecule is as high as possible. However, hole-anion pairs generated upon doping remain Coulombically bound to each other meaning that most charges are unable to contribute to transport.[14, 15] For instance, Pingel and Neher concluded that only 5% of charges contribute to transport in P3HT doped with the small-molecular dopant 2,3,5,6-tetrafluoro-7,7,8,8-tetracyanoquinodimethane (F$_4$TCNQ).[14] Several studies have recently shown that the size of the dopant molecule has a pronounced impact on the electrical conductivity of p-doped P3HT.[16, 17] Liang et al. found that the use of large molybdenum dithiolene complexes, which have a diameter of 11-14 Å, promotes delocalization of the polaron.[17] Further, Aubry et al. were able to effectively shield the hole polaron from its anion through the use of a large dodecaborane-based dopant with a diameter of 20 Å, which boosts the fraction of mobile charge carriers.[16]

In addition to ensuring a high number of mobile charges, it is important that the nanostructure of the polymer enables a high charge-carrier mobility. For molecularly doped P3HT the electrical conductivity strongly depends on the crystallinity of the polymer.[12, 18] Furthermore, uniaxial alignment has been widely explored as a tool to enhance the electrical conductivity and power factor in one direction. Qu et al. used directional epitaxial crystallization of P3HT with help of 1,3,5-trichlorobenzene, followed by doping with Fe(TFSI)$_3$ to reach a power factor of 38 µW m$^{-1}$ K$^{-2}$.[19] Further, some of us have used tensile drawing of bulk samples[20] or high-temperature rubbing of thin films[9, 21] to orient P3HT, achieving a power factor of 16, 21 and 56 µW m$^{-1}$ K$^{-2}$ upon subsequent doping with Mo(tfd-COCF$_3$)$_3$, FeCl$_3$, and F$_4$TCNQ, respectively (see Table 1).

**Table 1.** Selected literature values for the thermoelectric properties of P3HT at room temperature; [a]for oriented samples highest reported values measured along the direction of alignment are given.

| dopant | alignment[a] | $\sigma$ (S cm$^{-1}$) | $\alpha$ (µV K$^{-1}$) | $\alpha^2\sigma$ (µW m$^{-1}$ K$^{-2}$) | ref. |
|---|---|---|---|---|---|
| Fe(TFSI)$_3$ | no | 87 | 48 | 20 | 22 |
| Fe(TFSI)$_3$ | yes | 251 | 39 | 38 | 19 |
| F$_4$TCNQ | no | 12.7 | 46 | 2.7 | 12 |
| F$_4$TCNQ | no | 48 | 85 | 27 | 23 |
| F$_4$TCNQ | yes | 22 | 60 | 8.5 | 21 |
| F$_4$TCNQ | yes | 160 | 59 | 56 | 24 |
| Mo(tfd-COCF$_3$)$_3$ | yes | 12.7 | 112 | 16 | 20 |
| Mo(tfd)$_3$ + FeCl$_3$ | no | 68.5 | 8.1 | 4.7 | 17 |
| FeCl$_3$ | no | 42 | 105 | 46 | 6 |
| FeCl$_3$ | yes | 570 | 5.4 | 21 | 9 |
| FTS | no | 27.7 | 60 | 10 | 11 |

Here, we combine the aforementioned strategies with the aim of increasing the thermoelectric power factor that can be achieved with P3HT. To increase the order of the polymer, we chose to employ high-temperature rubbing, a technique that allows the preparation of highly aligned polymer films with an effective thickness of a few tens of nanometers. As the dopant we selected the molybdenum dithiolene complex Mo(tfd-COCF$_3$)$_3$ (see Figure 1 for chemical structure), which due to a high electron affinity EA ~ 5.6 eV[25, 26] offers a large driving force for oxidation of P3HT.

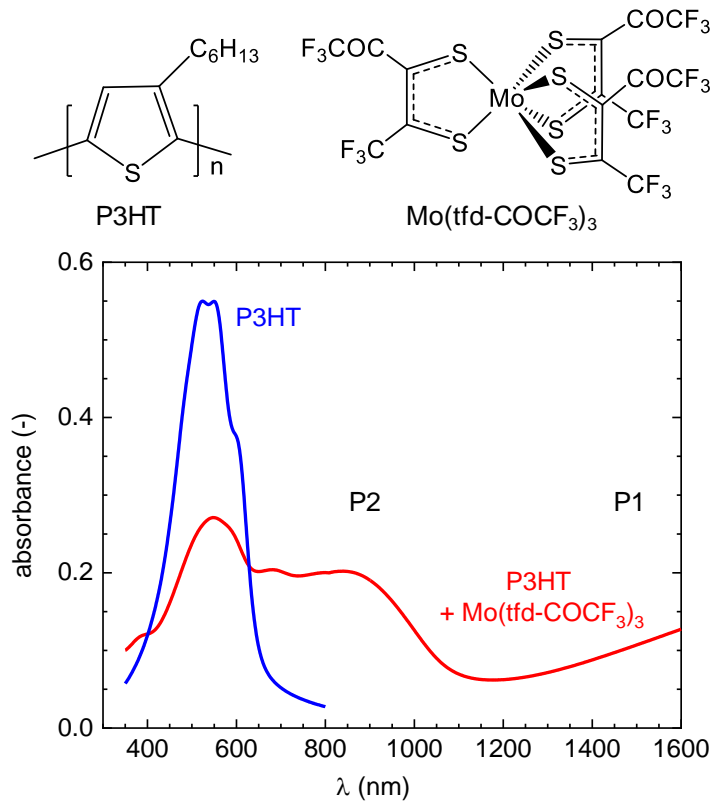

**Figure 1.** Chemical structure of P3HT and Mo(tfd-COCF$_3$)$_3$; UV-vis-NIR absorbance spectrum of a P3HT film before (blue) and after (red) sequential doping with Mo(tfd-COCF$_3$)$_3$ ($c_{Mo}$ ~ 1 g L$^{-1}$).

**Results and Discussion**

In a first set of experiments we studied doping of isotropic P3HT films. Sequential doping was carried out by drop-casting solutions of Mo(tfd-COCF$_3$)$_3$ in 1:1 acetonitrile:chloroform (AcN:CHCl$_3$) onto doctor-bladed P3HT films. We used an AcN:CHCl$_3$ solvent mixture because it resulted in a higher degree of doping than AcN solutions (CHCl$_3$ would dissolve P3HT). The dopant solution was allowed to remain on top of the film for 3 min, followed by spinning off excess solution (see Experimental for details). We recorded UV-vis-NIR spectra of films doped with solutions containing 0.1, 1 and 7.5 g L$^{-1}$ of the dopant. Doping gives rise to two pronounced absorption bands, P1 located in the infrared part of the spectrum and P2 centered at $\lambda_{P2}$ ~ 850 nm, while the absorption of the neat polymer is diminished in the doped samples (Figure 1 and Figure S1).

We were interested in estimating the number of generated charges, which (assuming that only polarons and no bipolarons are present) can be probed by examining the absolute absorbance of the polaronic peaks. Furthermore, neither neutral Mo(tfd-COCF$_3$)$_3$ nor its anion absorb at 800 nm.[27] A combination of spectroelectrochemistry and chronoamperometry allowed us to determine the molar attenuation coefficient $\varepsilon_{P2}$ of the first sub-bandgap polaron peak P2. Our electrochemical setup consisted of a spin-coated P3HT film on the indium tin oxide (ITO) working electrode, a platinum wire counter electrode and a silver wire as pseudo reference electrode, immersed in an electrolyte solution of 0.1M tetrabutylammonium hexafluorophosphate (TBAPF$_6$) in acetonitrile (AcN) (see Experimental for details). A cyclic voltammogram of P3HT indicates an oxidation onset of $E_{ox}$ ~ 0.55 V in our electrochemical setup (Figure 2a; note that the onset of P3HT vs ferrocene/ferrocenium is located at 0 V)[28]. We carried out a series of oxidation reactions at constant potentials between 0.55 and 0.75 V and recorded the transient current $I(t)$ (Figure 2b), which we integrated over time $t$ to obtain the total number of charges $Q$:

$$Q = \int_{t=0}^{\infty} I(t) dt \qquad (2)$$

Normalization by the sample volume in contact with the electrolyte yielded the charge density $Q_v$. At the end of each oxidation step we recorded a UV-vis-NIR spectrum and plotted the difference in absorbance $\Delta A$ between doped and undoped P3HT at 800 nm, normalized by the film thickness $d = 48$ nm (Figure 2c). A plot of $\Delta A/d$ vs. $Q_v$ shows a linear trend (Figure 2d), which indicates that only one absorbing species, i.e. polarons, are present up to a charge density of at least $Q_v = 3 \cdot 10^{26}$ m$^{-3}$. In agreement, Enengl et al. have shown that a significant amount of bipolarons only starts to form in P3HT at potentials larger than 0.5 V above the oxidation onset.[29] The slope of $\Delta A/d$ vs. $Q_v$ yields $\varepsilon_{P2} \sim (4.1 \pm 0.2) \cdot 10^3$ m$^2$ mol$^{-1}$ at 800 nm ($\varepsilon_{P2} \sim 41000$ M$^{-1}$ cm$^{-1}$). We would like to point out that cyclic voltammograms of P3HT can display several anodic waves that correspond to oxidation of ordered and disordered material.[28] Our analysis is limited to the first anodic wave where ordered P3HT is oxidized and hence the here reported value for $\varepsilon_{P2}$ corresponds to polarons in crystalline domains.

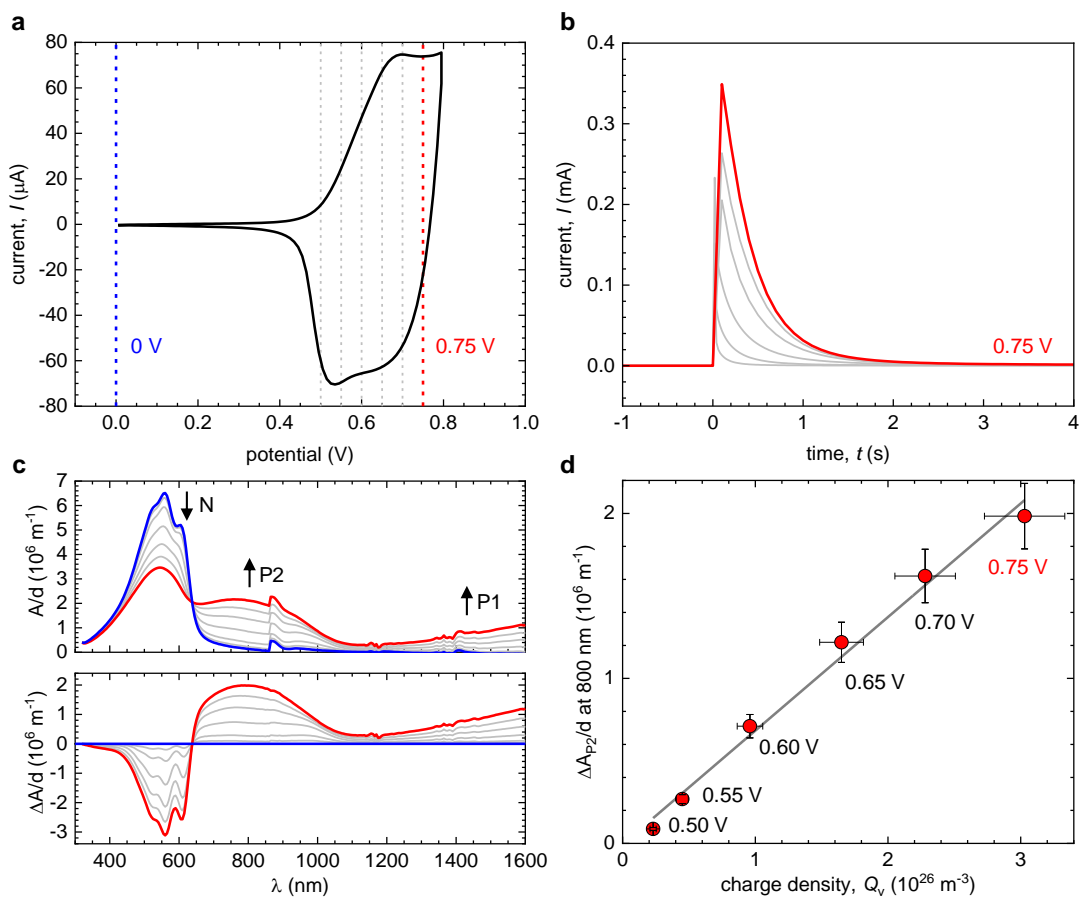

**Figure 2.** (a) Cyclic voltammogram of P3HT (P3HT has an oxidation onset around 0 V vs. ferrocene/ferrocenium according to ref. 28); (b) charging current from chronoamperometry, (c) UV-vis-NIR absorbance spectra from spectroelectrochemistry measurements recorded at different electrochemical potentials between 0 V (blue) and 0.75 V (red); thickness normalized absorbance $A/d$ (top) and difference in normalized absorbance $\Delta A/d$ between spectra of P3HT oxidized at different potentials and undoped P3HT at 0 V; (d) the differential absorbance $\Delta A_{P2}/d$ at 800 nm vs. charge density $Q_v$ from integration of the charging currents; the slope of the linear fit (gray) is the molar attenuation coefficient $\varepsilon_{P2}$.

The P2 polaron peak of electrochemically oxidized P3HT is centered at $\lambda_{P2} \sim 800$ nm, i.e. at a lower wavelength than chemically doped P3HT (cf. Figure 1). We explain this shift in $\lambda_{P2}$ with the difference in anion size. During electrochemical oxidation $PF_6^-$ counterions, which have a thermochemical radius of $r = 2.4$ Å,[30] enter the polymer from the electrolyte. Chemical doping instead produces Mo(tfd-COCF$_3$)$_3^-$ counterions, which have a much larger radius of $r$ ~7.5 Å (based on an estimate of the smallest sphere that could encompass the molecule/ion),[27] and hence results in a larger average polaron-anion distance.[31] Since the P2 polaron absorbance is broad we deem the difference in peak position to be minor. Hence, we used $\varepsilon_{P2}$ obtained from our electrochemistry experiments to estimate the charge-carrier density $N_v$ of chemically doped P3HT according to the Beer-Lambert law:

$$\Delta A_{P2} = \varepsilon_{P2} \cdot d \cdot N_v \qquad (3)$$

where $\Delta A_{P2}$ is the difference in absorbance at 800 nm between doped and undoped P3HT (Table 2). Mo(tfd-COCF$_3$)$_3$ can dope both ordered and disordered P3HT because of its high electron affinity of 5.6 eV. The here reported value for $\varepsilon_{P2}$ was measured by oxidizing ordered P3HT (cf. Figure 2), and may differ in case of disordered material, which would introduce an error in the here presented analysis of the charge-carrier density. In a recent study we used the same approach to estimate the charge-carrier density of chemically doped diketopyrrolopyrrole (DPP) and could confirm the value obtained for $N_v$ with electron paramagnetic resonance (EPR).[32] For samples sequentially doped with $c_{Mo}$ ~ 1 g L$^{-1}$ Mo(tfd-COCF$_3$)$_3$ in AcN:CHCl$_3$ we extract a value of $N_v$ ~ $(4.4 \pm 0.5) \cdot 10^{26}$ m$^{-3}$. P3HT has a density of about 1.1 g cm$^{-3}$,[33] and hence contains $4 \cdot 10^{27}$ thiophene rings per m$^3$, which implies an oxidation level of 11 %, i.e. one polaron for every ten thiophene repeat units. The same samples display an electrical conductivity of $\sigma$ ~ $(260 \pm 26)$ S cm$^{-1}$, which according to equation 1 translates to a charge-carrier mobility of $\mu$ ~ $(3.3 \pm 0.5)$ cm$^2$ V$^{-1}$ s$^{-1}$ (Table 2). Note that this value is only slightly higher than the 1 cm$^2$ V$^{-1}$ s$^{-1}$ reported for P3HT doped with F4TCNQ.[34] Furthermore, we measured a high Seebeck coefficient of $\alpha$ ~ 44 µV K$^{-1}$, yielding a thermoelectric power factor of $\alpha^2\sigma$ ~ $(50 \pm 10)$ µW m$^{-1}$ K$^{-2}$, which is comparable to the highest values reported for isotropic P3HT (cf. Table 1).

**Table 2.** Electrical conductivity $\sigma$, Seebeck coefficient $\alpha$ (error $\pm$ 2 µV K$^{-1}$), power factor $\alpha^2\sigma$, density and mobility of charges, $N_v$ and $\mu$, and oxidation level for doctor-bladed films with a thickness $d$ sequentially doped with an AcN:CHCl$_3$ solution containing a concentration of $c_{Mo}$ of Mo(tfd-COCF$_3$)$_3$; $\sigma$ and $\alpha$ were measured immediately after doping but remained relatively stable during storage in a glove box for one week (Figure S2).

| $c_{Mo}$ (g L$^{-1}$) | $d$ (nm) | $N_v$ (10$^{26}$ m$^{-3}$) | ox. level (wt%) | $\sigma$ (S cm$^{-1}$) | $\alpha$ (µV K$^{-1}$) | $\alpha^2\sigma$ (µW m$^{-1}$ K$^{-2}$) | $\mu$ (cm$^2$ V$^{-1}$ s$^{-1}$) |
|---|---|---|---|---|---|---|---|
| 0.1 | 55 ± 6 | 0.9 ± 0.1 | 2 | 1.1 ± 0.1 | 139 | 2.0 ± 0.3 | 0.07 ± 0.01 |
| 1 | 57 ± 6 | 4.4 ± 0.5 | 11 | 260 ± 26 | 44 | 50 ± 10 | 3.3 ± 0.5 |
| 7.5 | 51 ± 5 | 5.0 ± 0.6 | 13 | 285 ± 29 | 35 | 35 ± 8 | 3.2 ± 0.5 |

In a further set of experiments, we doped a series of thin-film samples that featured a high degree of in-plane alignment. Oriented samples were prepared by high-temperature rubbing at 186 °C (see Experimental section and refs. 21, 35, 36 for details). We characterized the degree of alignment with polarized optical microscopy (Figure 3a) and polarized UV-vis-NIR spectroscopy (Figure 3b and Figure S3). The absorption of P3HT is considerably stronger when measured parallel to the rubbing direction, with a maximum dichroic ratio of $A_\parallel/A_\perp \sim 11.5$ at 633 nm, measured for rubbed P3HT (Figure S4), which confirms uniaxial alignment of the conjugated backbone. This is also supported by electron diffraction, which shows that the 002 reflection related to the monomer repeat periodicity ($q_{002} \sim 1.63$ Å$^{-1}$) is located along the meridian, i.e. along the rubbing direction R (Figure 3c). Instead, both the *h*00 reflections (*h* = 1-3; $q_{100} \sim 0.379$ Å$^{-1}$) associated with lamellar stacking of P3HT and the 020 reflection ($q_{020} \sim 1.67$ Å$^{-1}$) from π-stacking are oriented along the equator (Figure 3c). The simultaneous presence of both *h*00 and 020 reflections is characteristic for a mixture of face-on and edge-on crystalline domains within the polymer films.

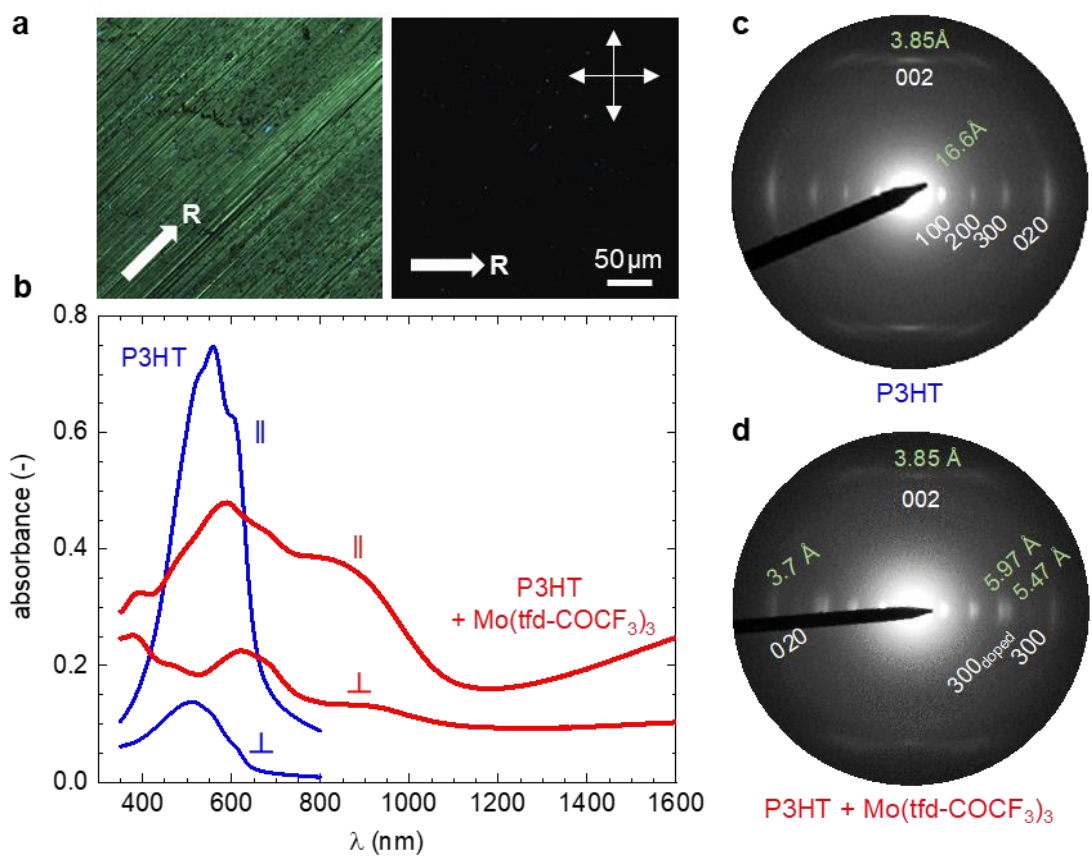

**Figure 3.** (a) Polarized optical microscopy images of rubbed P3HT films sequentially doped with $c_{Mo} \sim 1$ g L$^{-1}$; crossed double arrows indicate the orientation of polarizer and analyzer; white arrow indicates the rubbing direction R; (b) polarized UV-vis-NIR absorbance spectra of rubbed P3HT and the same P3HT film doped with $c_{Mo} \sim 1$ g L$^{-1}$ measured parallel (∥) and perpendicular (⊥) to the rubbing direction R; (c, d) electron diffraction patterns of neat rubbed P3HT and rubbed P3HT doped with Mo(tfd-COCF$_3$)$_3$.

Doping of rubbed films was again done by sequential doping with Mo(tfd-COCF$_3$)$_3$ dissolved in AcN:CHCl$_3$. We note that the positions of the *h*00 reflections (*h* = 1-3) split upon doping with Mo(tfd-COCF$_3$)$_3$ (Figure 3d). For instance, we observe two 300 reflections, one close to the original position at $q_{300}$ ~ 1.15 Å$^{-1}$ and one at $q_{300}$ ~ 1.05 Å$^{-1}$. The 020 peak related to π-stacking does not split but slightly shifts to $q_{020}$ ~ 1.70 Å$^{-1}$ whereas the 002 reflection is unchanged. These observations indicate that the films contain both non-doped and doped P3HT crystallites, with the dopant intercalated in the side chain layers as suggested by the expansion of the lattice along the side chain direction.

Interestingly, rinsing of as-doped samples with AcN reverts the diffraction pattern to that of neat P3HT (Figure S5). Rinsing reduces the oxidation level, indicated by an increase in the neutral polymer absorbance (Figure S6), and leads to a threefold reduction in electrical conductivity. Analysis of the P2 absorbance band indicates a slight reduction in charge-carrier density to $N_v$ ~ (3.3 ± 0.4)·10$^{26}$ m$^{-3}$. We argue that the dopant that remains after rinsing is only located in amorphous regions.

Polarized UV-vis-NIR spectra reveal a significant reduction of the neutral polymer absorption independent of the dopant concentration ($c_{Mo}$ ~ 1, 2.5 and 7.5 g L$^{-1}$), indicating that all samples were strongly doped. The two polaron absorbance peaks in the near and far-IR display considerable anisotropy with the stronger absorbance when measured parallel to the rubbing direction (cf. Figure 3b). We argue that polarons delocalize along the oriented polymer backbone as recently confirmed by Ghosh et al.[37] We again used the molar attenuation coefficient $\varepsilon_{P2}$ to estimate the number of charge carriers. To be able to compare the polarized absorbance of the rubbed samples with the isotropic absorbance of electrochemically oxidized P3HT, we used the average absorbance $\Delta A_{P2} = (\Delta A_\parallel + \Delta A_\perp)/2$ at 800 nm (cf. Fig. 3b). From this value we subtracted $A_\perp$ of rubbed and doped films at 1200 nm to account for the apparent vertical offset of the UV-vis-NIR spectra, which we tentatively assign to light scattering by excess dopant on top of the film (cf. Figure S6). The deduced charge-carrier density of rubbed films is comparable to values extracted for isotropic samples (Table 3). For rubbed P3HT films doped with $c_{Mo} \sim$ 1 g L$^{-1}$ Mo(tfd-COCF$_3$)$_3$ we obtain $N_v \sim (4.0 \pm 0.4) \cdot 10^{26}$ m$^{-3}$, which corresponds to an oxidation level of 10 %.

The electrical conductivity is considerably higher along the rubbing direction, e.g. $\sigma_\parallel \sim (509 \pm 51)$ S cm$^{-1}$ versus $\sigma_\perp \sim (84 \pm 8)$ S cm$^{-1}$ for P3HT doped with $c_{Mo} \sim$ 1 g L$^{-1}$ Mo(tfd-COCF$_3$)$_3$ (Table 3), which gives rise to an anisotropy of $\sigma_\parallel/\sigma_\perp \sim 6$ (see Figure S2 for stability of conductivity over time). Together with our estimate for $N_v$ we calculate a charge-carrier mobility of $\mu_\parallel \sim (7.1 \pm 0.1)$ cm$^2$ V$^{-1}$ s$^{-1}$ and $\mu_\perp \sim (1.2 \pm 0.2)$ cm$^2$ V$^{-1}$ s$^{-1}$, which represent average values for all mobile plus bound charges. We note that $\mu_\parallel$ is significantly higher than values measured for isotropic samples (cf. Table 2), while $\mu_\perp$ is lower.

**Table 3.** Electrical conductivity $\sigma$, Seebeck coefficient $\alpha$ (error $\pm$ 2 µV K$^{-1}$), power factor $\alpha^2\sigma$, density and mobility of charges, $N_v$ and $\mu$, and oxidation level for rubbed films with a thickness $d$ sequentially doped with an AcN:CHCl$_3$ solution containing a concentration of $c_{Mo}$ of Mo(tfd-COCF$_3$)$_3$; subscripts ∥ and ⊥ refer to in-plane values measured parallel and perpendicular to the rubbing direction; $\sigma$ and $\alpha$ were measured immediately after doping but remained relatively stable during storage in a glove box for one week, and even after subsequent exposure to air (Figure S2).

| $c_{Mo}$ (g L$^{-1}$) | $d$ (nm) | $N_v$ (10$^{26}$ m$^{-3}$) | ox. level (%) | $\sigma_\parallel$ (S cm$^{-1}$) | $\sigma_\perp$ (S cm$^{-1}$) | $\alpha_\parallel$ (µV K$^{-1}$) | $\alpha_\perp$ (µV K$^{-1}$) | $\alpha_\parallel^2\sigma_\parallel$ (µW m$^{-1}$ K$^{-2}$) | $\alpha_\perp^2\sigma_\perp$ (µW m$^{-1}$ K$^{-2}$) | $\mu_\parallel$ (cm$^2$ V$^{-1}$ s$^{-1}$) | $\mu_\perp$ (cm$^2$ V$^{-1}$ s$^{-1}$) |
|---|---|---|---|---|---|---|---|---|---|---|---|
| 1 | 44 | 4.0 ± 0.4 | 10 | 509 ± 51 | 84 ± 8 | 56 | 13 | 160 ± 27 | 1.0 ± 0.5 | 7.1 ± 1.1 | 1.2 ± 0.2 |
| 2.5 | 58 | 3.6 ± 0.4 | 9 | 593 ± 59 | 116 ± 12 | 47 | 17 | 131 ± 12 | 3.0 ± 0.8 | 9.3 ± 1.4 | 1.8 ± 0.3 |
| 7.5 | 53 | 4.4 ± 0.5 | 11 | 681 ± 68 | 50 ± 5 | 43 | 6 | 126 ± 12 | 0.2 ± 0.1 | 8.7 ± 1.3 | 0.6 ± 0.1 |

In agreement with previous studies,[9, 21, 24] the Seebeck coefficient also displays a high degree of in-plane anisotropy, e.g. $\alpha_\parallel/\alpha_\perp \sim 4$ in case of P3HT doped with $c_{Mo} \sim 1$ g L$^{-1}$ Mo(tfd-COCF$_3$)$_3$. An absolute Seebeck coefficient of $\alpha_\parallel \sim (56 \pm 2)$ µV K$^{-1}$ gives rise to a maximum power factor of $\alpha_\parallel^2 \sigma_\parallel \sim (160 \pm 27)$ µW m$^{-1}$ K$^{-2}$ (Table 3), which is the highest value so far reported for P3HT (cf. Table 1). We note that $\alpha_\parallel^2 \sigma_\parallel$ slightly exceeds the empirical trend $\alpha^2 \sigma \propto \sqrt{\sigma}$ that is often observed for isotropic samples,[11-13] while $\alpha_\perp^2 \sigma_\perp$ falls short of the predicted value (Figure S7).

The anisotropy of the Seebeck coefficient and conductivity was recently studied using kinetic Monte Carlo (kMC) simulations to describe thermoelectric measurements on doped poly(2,5-bis(3-dodecyl-2-thienyl)thieno[3,2-*b*]thiophene) (PBTTT).[38] In brief, these kMC simulations account for variable-range hopping on a regular lattice. Structural anisotropy is reflected in the model by an increased delocalization in the parallel direction compared to the perpendicular direction, i.e. $\xi_\parallel > \xi_\perp$ where $\xi$ is the localization length (cf. ref. 38 for details). This model can reproduce the trend of the experimental data in the parallel and perpendicular direction measured for rubbed P3HT doped with Mo(tfd-COCF$_3$)$_3$, when assuming an anisotropy ratio of $\xi_\parallel/\xi_\perp = 4$ and an attempt-to-hop frequency of $\nu_0 = 2\cdot 10^{13}$ s$^{-1}$ (cf. Figure 4). We argue that transport in the highly doped samples studied here is dominated by hopping between discrete sites along the direction of orientation, while inter-chain transport only plays a minor role.[18, 38] In a system with such a distinct orientation, carriers are forced to hop along the parallel direction and therefore have a limited ability to optimize their path with respect to energy. This results in an upward shift of the transport energy $E_{tr}$ in the parallel direction and as $\alpha \propto (E_F - E_{tr})/T$ with $E_F$ the Fermi energy also in an increase of $\alpha_\parallel$ compared to $\alpha_\perp$.[38]

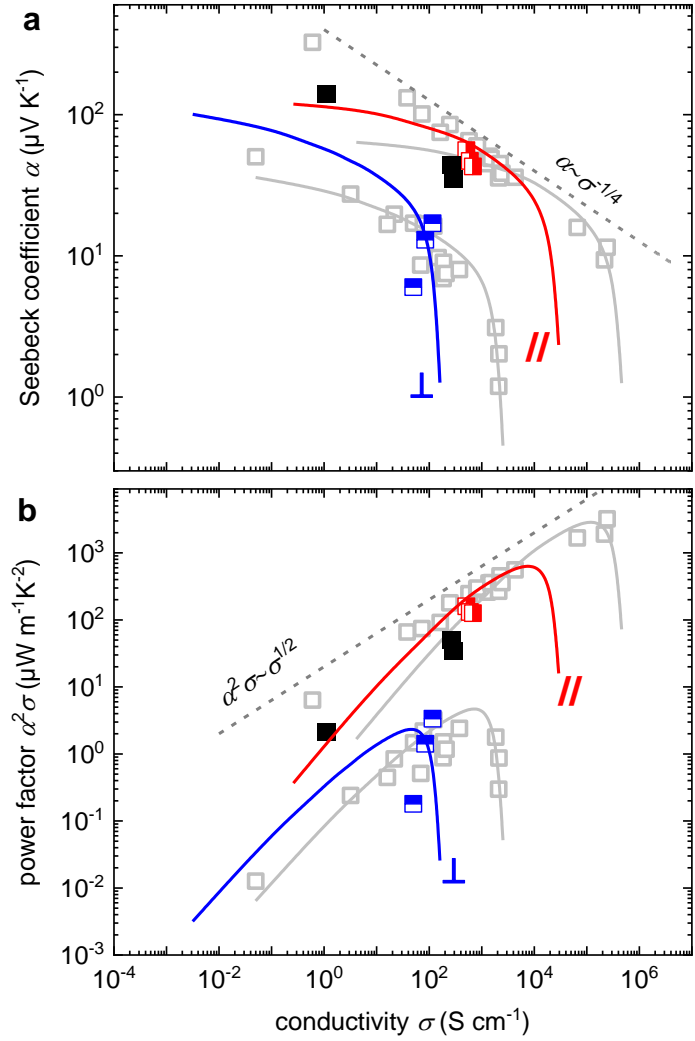

**Figure 4.** (a) Seebeck coefficient $\alpha$ and (b) power factor $\alpha^2\sigma$ versus electrical conductivity $\sigma$ of isotropic P3HT (black squares), and rubbed P3HT measured parallel (red squares and line) and perpendicular to the rubbing direction (blue squares and line); grey data points are literature values extracted from refs. 9, [38] for PBTTT. Full lines correspond to simulations calculated from the kMC model, with an anisotropy ratio of $\xi_\parallel/\xi_\perp = 4$, where $\xi$ is the localization length in the parallel and perpendicular direction, respectively. The attempt to hop frequency was set to $\nu_0 = 2\cdot 10^{13}$ s$^{-1}$ and the thermopower was rescaled by a factor of 0.1 in both directions. ∥ and ⊥ refer to the parallel and perpendicular direction, respectively. The dashed gray line shows the empirical trends $\alpha \propto \sigma^{-0.25}$ and $\alpha^2\sigma \propto \sqrt{\sigma}$.

**Conclusions**

We conclude that the combination of a large dopant such as Mo(tfd-COCF$_3$)$_3$ in combination with structural anisotropy, here achieved by high-temperature rubbing, is a powerful means to improve the thermoelectric properties of conjugated polymers. A high conductivity of $\sigma_\parallel \sim (509 \pm 51)$ S cm$^{-1}$ is obtained thanks to a high charge-carrier mobility along the rubbing direction. At the same time, the Seebeck coefficient remains high with $\alpha_\parallel \sim (56 \pm 2)$ µV K$^{-1}$, and as a result we achieve a record thermoelectric power factor of $\alpha_\parallel^2 \sigma_\parallel \sim (160 \pm 27)$ µW m$^{-1}$ K$^{-2}$ for P3HT.

**Acknowledgements**

J.H., A.I.H. and C.M. gratefully acknowledge financial support from the Swedish Research Council through grant no. 2016-06146 and 2018-03824, the Knut and Alice Wallenberg Foundation through a Wallenberg Academy Fellowship, and the European Research Council (ERC) under grant agreement no. 637624. D.S. acknowledges funding from the European Union's Horizon 2020 research and innovation program under the Marie Skłodowska-Curie Grant Agreement No. 799477. S.R.M., S.B., and Y.Z. thank the U. S. National Science Foundation for support of this work through the DMREF program, under award No. DMR-1729737. V.U. acknowledges financial support from the IRTG Softmatter project. M.B. thanks the Agence Nationale de la Recherche for financial support through project ANR -17-CE05-0012.

**Supporting Information**

UV-vis-NIR spectra of neat and doped P3HT films; electrical conductivity vs. time; dichroic ratio of rubbed P3HT; electron diffraction patterns of rubbed and doped P3HT before and after rinsing with acetonitrile; plots of Seebeck coefficient and thermoelectric power factor vs. electrical conductivity; UV-vis absorbance spectra vs. film thickness.

**Experimental**

**Materials.** P3HT was purchased from Merck ($M_n$ = 24 kg mol$^{-1}$, PDI = 1.8, regioregularity = 95.9 %). Sodium polystyrene sulfonate (NaPSS) was purchased from Sigma Aldrich. Mo(tfd-COCF$_3$)$_3$ was synthesized as previously described.[26] Anhydrous acetonitrile (AcN; purity > 99.9%), chloroform (CHCl$_3$; purity > 99.5%), *ortho*-dichlorobenzene (oDCB; purity > 99%) and TBAPF$_6$ were purchased from Sigma-Aldrich and used as received.

**Orientation and doping of thin films.** P3HT films were prepared by doctor-blading a hot solution of P3HT dissolved in oDCB (10 g L$^{-1}$) at 165 °C on glass substrates coated with a thin layer of sodium polystyrene sulfonate (NaPSS; spin-coated from 10 g L$^{-1}$ aqueous solution). The orientation of thin P3HT films by high temperature rubbing followed the protocol described in refs. [21, 35, 36], and was performed using a home-built rubbing machine consisting of a translating hot plate at 186 °C, on which the sample is placed, and a rotating cylinder covered with a microfiber cloth. The effective thickness was estimated by comparing the UV-vis absorbance of annealed films (315 °C for 5 min) at 557 nm with the calibration curve provided in SI Figure S8. All doping experiments were carried out under inert atmosphere in a glovebox from Jacomex. Dopant solutions were prepared by mixing equal amounts of AcN and CHCl$_3$, followed by addition of Mo(tfd-COCF$_3$)$_3$ powder yielding a concentration of 7.5 g L$^{-1}$. Lower concentrations (0.1; 1; 2.5 g L$^{-1}$) were obtained by dilution of the 7.5g L$^{-1}$ stock solution with 1:1 AcN:CHCl$_3$. Doping was performed by drop-casting the dopant solution on top of P3HT films, which was left in contact for 3 min before spinning off the solution. Sequential doping did not alter the film thickness as indicated by atomic force microscopy (AFM) of a spin-coated P3HT film before and after doping (30 vs. 31 nm), using a Digital Instruments Nanoscope IIIA.

**UV-vis-NIR spectroscopy.** UV-vis-NIR absorbance spectra of chemically doped samples were recorded with a Varian Cary5000 spectrometer. Polarized incident light was used in case of rubbed thin films. The spectral resolution was 1 nm.

**Spectroelectrochemistry.** For the spectroelectrochemical measurements thin P3HT films were cast onto ITO coated glass (width 6 mm) via spin-coating from a solution of 10 g L$^{-1}$ P3HT in oDCB. The spectroelectrochemical setup consisted of a 1cm x 1cm UV-vis quartz cuvette with a custom-made Teflon lid, which holds a 3-electrode setup comprising the ITO/polymer sample as working electrode, a platinum wire counter electrode and a silver wire as pseudo reference electrode. All electrochemical measurements were performed in a solution of 0.1M TBAPF$_6$ in dry and degassed acetonitrile; the potential scale of our setup lies at 0.4 to 0.5 V relative to ferrocene/ferrocenium. Cyclic voltammetry was performed at 100 mV s$^{-1}$. For chronoamperometry measurements the film was first de-doped upon application of a potential of 0 V for 60 s, before the respective positive oxidation potential between 0.55 V and 075 V was applied for 300 s. UV-vis-NIR absorption spectra were recorded using a PerkinElmer Lambda 1050 spectrophotometer after 120 s once the electrochemical current had stabilized. The amount of injected charges was calculated by integrating the electrochemical current over time, using the background current as the baseline. The charge-carrier density was calculated by diving the amount of the injected charge carriers by the film volume in contact with the electrolyte. The film thickness was estimated by comparing the peak absorbance of undoped P3HT at 557 nm with the calibration curve provided in Figure S8.

**Transmission electron microscopy (TEM).** Samples were prepared by first depositing a thin amorphous carbon layer on top of pieces of oriented polymer films on NaPSS/glass substrates using an auto 306 Edwards evaporator. Then, oriented areas were identified with an optical microscope (Leica DMR-X) and floated off on distilled water, which dissolved the sacrificial NaPSS layer. The carbon-coated P3HT films were recovered on TEM copper grids. Finally, grids were immediately doped by dropping 10 µL of the dopant solution onto the samples, which dried on the sample through evaporation of the solvent. TEM was performed in bright field and diffraction mode using a CM12 Philips microscope equipped with a MVIII (Soft Imaging System) charge coupled device camera. To avoid dedoping under the electron beam, exposure was set to a minimum using the low dose system.

**Thermoelectric measurements.** Samples were fabricated using the procedure described in refs. 8, 21. Gold contacts (40 nm thick) in a four-point probe geometry (1 mm spacing between electrodes, 5 mm length) were deposited on glass substrates via controlled thermal evaporation through a shadow mask. A first layer of chromium (2.5 nm thick) was deposited prior the gold to promote adhesion. The geometry of deposited gold electrodes allowed determination of electrical resistivity and thermopower for the same sample both parallel and perpendicular to the rubbing direction. Oriented films of P3HT on NaPSS/glass were floated off with distilled water, recovered with glass substrates on which gold electrodes had been deposited, and finally doped in a glovebox (Jacomex). Thermoelectrical characterization was performed immediately after doping. A Keithley 4200-SCS and a Lab Assistant Semiprobe station in the glovebox under nitrogen atmosphere were used to determine the sheet resistance $R$ (using a four-point probe geometry) and thermopower, following the procedure described in refs. 8, 21. The resistivity $\rho$ was obtained according to $\rho = 1.81 \cdot R \cdot t$, where $t$ is the film thickness (determination of the geometrical correction factor 1.81 is described in ref. 21). Reported conductivity values are the average of two to four samples. The thermopower was measured via the differential temperature method as described in ref. 21, by establishing a temperature gradient across the sample either along or perpendicular to the rubbing direction.